\documentclass[conference, final, letterpaper]{IEEEtran}

\usepackage[utf8]{inputenc}
\usepackage{graphicx}
\usepackage{stfloats}
\usepackage{amssymb}
\usepackage[cmex10]{amsmath}
\usepackage{amsthm} 
\usepackage{array}
\usepackage{pifont}
\usepackage[usenames,dvipsnames]{xcolor}
\usepackage{hyperref}
\hypersetup{colorlinks=false, pdfborderstyle={/S/U/W 1},pdfborder=0 0 1, citebordercolor=Blue, filebordercolor=Red, linkbordercolor=Red, urlbordercolor=Blue}
\usepackage{url}

\usepackage[backend=bibtex8, style=ieee-alex, url=false, isbn=false, doi=true, maxnames=5, firstinits=true]{biblatex}[2012/12/20 v2.4]
\usepackage[english]{babel}
\usepackage[english = american]{csquotes}
\MakeOuterQuote{"}
\DeclareListFormat{language}{}
\bibliography{../../HTGen}

\usepackage[ruled,vlined,titlenumbered]{algorithm2e}
\usepackage{tabularx}
\newtheorem{definition}{Definition}
\newtheorem{theorem}{Theorem}

\newtheorem{proposition}{Proposition}

\newtheorem{example}{Example}
\newtheorem{corollary}{Corollary}
\DeclareMathOperator{\lcm}{lcm}
\DeclareMathOperator{\diag}{diag}
\newcommand{\gcdab}[2]{\ensuremath{\gcd(#1,#2)}}
\newcommand{\lcmab}[2]{\ensuremath{\lcm(#1,#2)}}
\DeclareMathOperator{\defi}{def}
\newcommand{\defeq}{\overset{\defi}{=}}
\newcommand{\defequiv}{\overset{\defi}{\equiv}}
\DeclareMathOperator{\rank}{rank}

\newcommand{\F}[1]{\mathbb F_{#1}}
\newcommand{\Fq}{\F{q}}
\newcommand{\Fxsub}[1]{\ensuremath{\mathbb{F}_{#1}[X]}}
\newcommand{\Fqx}{\Fxsub{q}}

\newcommand{\Z}{\ensuremath{\mathbb{Z}}}
\renewcommand{\vec}[1]{\mathbf #1}
\newcommand{\M}[2][\empty]{
  \ifthenelse{\equal{#1}{\empty}}
    {\ensuremath{\mathbf{#2}}}
    {\ensuremath{{\mathbf{#2}}_{#1}}}
}
\newcommand{\SET}[1]{\ensuremath{\mathsf{#1}}}

\renewcommand{\b}{\mathbf b}

\newcommand{\LIN}[3]{\ensuremath{[#1,#2]_{#3}}}
\newcommand{\LINq}[2]{\ensuremath{[#1,#2]_{q}}}
\newcommand{\defset}[2][\empty]{
  \ifthenelse{\equal{#1}{\empty}}
    {\ensuremath{\SET{D}_{#2}}}
    {\ensuremath{\SET{D}^{[#1]}_{#2}}}
}
\newcommand{\CYCa}{\ensuremath{\mathcal{A}}}
\newcommand{\CYCan}{\ensuremath{n_1}} 
\newcommand{\CYCak}{\ensuremath{k_1}} 
\newcommand{\CYCad}{\ensuremath{d_1}} 
\newcommand{\CYCas}{\ensuremath{s_1}} 
\newcommand{\CYCb}{\ensuremath{\mathcal{B}}}
\newcommand{\CYCbn}{\ensuremath{n_2}} 
\newcommand{\CYCbk}{\ensuremath{k_2}} 
\newcommand{\CYCbd}{\ensuremath{d_2}} 
\newcommand{\CYCbs}{\ensuremath{s_2}} 
\newcommand{\CYC}{\ensuremath{\mathcal{C}}}
\newcommand{\CYCn}{\ensuremath{n}} 
\newcommand{\CYCk}{\ensuremath{k}} 
\newcommand{\CYCd}{\ensuremath{d}} 
\newcommand{\CYCs}{\ensuremath{s}} 
\newcommand{\inta}{\ensuremath{a}}
\newcommand{\intb}{\ensuremath{b}}
\newcommand{\HTconst}{\ensuremath{f}}
\newcommand{\HTconsta}{\ensuremath{f_1}}
\newcommand{\HTconstb}{\ensuremath{f_2}}
\newcommand{\lenseq}{\ensuremath{\delta}}
\newcommand{\noseq}{\ensuremath{\nu}}
\newcommand{\HTmult}{\ensuremath{m}}
\newcommand{\HTmulta}{\ensuremath{m_1}}
\newcommand{\HTmultb}{\ensuremath{m_2}}

\renewcommand{\tilde}{\widetilde}

\begin{document}
\title{Generalizing Bounds on the Minimum Distance of Cyclic Codes Using Cyclic Product Codes}

\IEEEoverridecommandlockouts

\author{\IEEEauthorblockN{Alexander Zeh and Antonia Wachter-Zeh}\thanks{This work has been supported by DFG,
Germany, under grants BO~867/22-1 and BO~867/21-1.}
\IEEEauthorblockA{Institute of Communications Engineering\\
Ulm University, Ulm, Germany\\
\texttt{\{alex, antonia\}@} \\
\texttt{codingtheory.eu}}
\and
\IEEEauthorblockN{Maximilien Gadouleau}
\IEEEauthorblockA{School of Engineering \& \\ Computing Sciences (ECS)\\
Durham University, Durham, UK\\
\texttt{m.r.gadouleau@durham.ac.uk}}
\and
\IEEEauthorblockN{Sergey Bezzateev}
\IEEEauthorblockA{Saint Petersburg State University\\ of Airspace Instrumentation\\
St. Petersburg, Russia\\
\texttt{bsv@aanet.ru}}
}

\maketitle

\begin{abstract}
Two generalizations of the Hartmann--Tzeng (HT) bound on the minimum distance of $q$-ary cyclic codes are proposed. The first one is proven by embedding the given cyclic code into a cyclic product code. Furthermore, we show that unique decoding up to this bound is always possible and outline a quadratic-time syndrome-based error decoding algorithm. The second bound is stronger and the proof is more involved.

Our technique of embedding the code into a cyclic product code can be applied to other bounds, too and therefore generalizes them.
\end{abstract}


\section{Introduction} 
Cyclic codes play a central role in (distributed) storage and communication systems. However, determining their minimum distance from a given defining set is an open research problem. Many lower bounds on the minimum distance and efficient decoding algorithms up to these bounds exist.

In the 1970s, Hartmann and Tzeng (HT,~\cite{hartmann_decoding_1972, hartmann_generalizations_1972}) generalized  the well-known bound by Bose, Ray-Chaudhuri~\cite{bose_class_1960} and Hocquenghem~\cite{hocquenghem_codes_1959} (BCH). Feng and Tzeng~\cite{feng_generalized_1989,feng_generalization_1991} extended the BCH decoding algorithms~\cite{massey_shift-register_1969, sugiyama_method_1975} to decode in quadratic-time up to the HT bound. Further extensions of the BCH bound were \textit{inter alia} developed by Roos~\cite{roos_generalization_1982,roos_new_1983}, van Lint and Wilson~\cite{van_lint_minimum_1986} (denoted as AB or Shifting method), Duursma and Kötter~\cite{duursma_error-locating_1994}, Boston~\cite{boston_bounding_2001}, Duursma and Pellikaan~\cite{duursma_symmetric_2006} and Betti and Sala~\cite{betti_new_2006}.

Our first generalization of the HT bound uses the idea of cyclic product codes (see~\cite{burton_cyclic_1965, abramson_cascade_1968, lin_further_1970} and Ch. 18 in~\cite{macwilliams_theory_1988}) and can be applied to other bounds, too.
The second approach also associates another cyclic code, but the direct connection to cyclic product codes is not clear.

In contrast to our previous contributions~\cite{zeh_decoding_2012, zeh_new_2012}, we provide a generalization of the HT bound and show proofs of the statements by means of cyclic product codes.


Our contribution is structured as follows. In Section~\ref{sec_CyclicProduct}, we give necessary preliminaries on cyclic codes, recall the HT bound~\cite{hartmann_decoding_1972, hartmann_generalizations_1972} and provide basic properties of cyclic product codes as they were described first in~\cite{burton_cyclic_1965}. 
The first generalization of the HT bound is proven in Section~\ref{sec_GenHTBoundProd} and the second one in Section~\ref{sec_GenHTBoundFrac}. The syndrome-based decoding approach up to the first bound is described in Section~\ref{sec_DecGenHTBound}. Section~\ref{sec_Conclusion} concludes our paper.

%
%
%
%
%

\section{Cyclic Codes and Cyclic Product Codes} \label{sec_CyclicProduct}
\subsection{Notation}
Let $\Z$ denote the set of integers, $\Fq$ the finite field of order $q$ and $\Fqx$ the polynomial ring over $\Fq$ with indeterminate $X$. 
A vector of length $n$ is denoted by a lowercase bold letter as $\vec{v} = (v_0 \, v_1 \, \dots \, v_{n-1})$.
An $m \times n$ matrix is denoted by a capital bold letter as $\M{M}=\| m_{i,j}\|_{i=0,j=0}^{m-1,n-1}$. 
A set is denoted by a capital letter sans serif like $\SET{D}$.

A linear $\LINq{\CYCn}{\CYCk}$ code of length $\CYCn$ and dimension $\CYCk$ over $\Fq$ is denoted by a calligraphic letter like $\CYC$ and its minimum Hamming distance by $\CYCd$.

\subsection{Cyclic Codes}
An \LINq{n}{k} cyclic code $\CYC$ with distance $d$ is an ideal in the ring $\Fqx / (X^n-1)$ generated by $g(X)$. 
The generator polynomial $g(X)$ has roots in the splitting field $\F{q^s}$, where $n
\mid (q^s -1)$. 
The primitive element of order $\CYCn$ is $\alpha$ and the defining set $\defset{\CYC}$ of an \LINq{\CYCn}{\CYCk} cyclic code $\CYC$ is:
\begin{equation} \label{eq_definingset}
\defset{\CYC}  =  \big\{ 0 \leq i \leq n-1 \, | \, g(\alpha^i)=0 \big\} .
\end{equation}
Furthermore, we introduce the following short-hand notations for a given $z \in \Z$:
\begin{equation} \label{eq_definingsetext}
\begin{split}
\defset[z,\otimes]{\CYC} & \defeq   \big\{ (i \cdot z) \bmod \CYCn  \; | \; i \in \defset{\CYC} \big\},\\
\defset[z,+]{\CYC} & \defeq  \big\{ (i+z) \; | \; i \in \defset{\CYC} \big\}.
\end{split}
\end{equation}
Let us recall the HT bound \cite{hartmann_decoding_1972, hartmann_generalizations_1972} and present it in polynomial form, which we use later on.
\begin{theorem}[HT Bound~\cite{hartmann_decoding_1972, hartmann_generalizations_1972}] \label{theo_HTBound}
Let an $\LINq{\CYCn}{\CYCk}$ cyclic code $\CYC$ with minimum distance $\CYCd$ be given. Let $\alpha$ denotes an element of order $n$.
Let four integers $\HTconst$, $\HTmult$, $\lenseq$ and $\noseq$ with $\HTmult \neq 0$ and $\gcd(n,\HTmult) = 1$, $\lenseq \geq 2$ and $\noseq \geq 0$ be given, such that:

\begin{equation} \label{eq_HTBound}
\sum_{i=0}^{\infty} c(\alpha^{\HTconst+i\HTmult+j})X^i \equiv 0 \mod X^{\lenseq-1}, \quad \forall j=0,\dots,\noseq,
\end{equation}
holds for all $c(x) \in \CYC$ and some integers $\lenseq \geq 2$ and $\noseq \geq 0$.

Then, $d \geq \lenseq+\noseq$.
\end{theorem}
Note that for $\noseq = 0$, the HT bound becomes the BCH bound~\cite{hocquenghem_codes_1959, bose_class_1960}.\\
Let $\CYCa$ and $\CYCb$ be \LIN{\CYCan}{\CYCak}{q} and \LIN{\CYCbn}{\CYCbk}{q} linear codes over $\Fq$ with minimum Hamming distance $\CYCad$ and $\CYCbd$. For simplicity, we assume that the first $\CYCak |\CYCbk$ symbols are the information symbols of $\CYCa | \CYCb$.
\begin{definition}[Product Code] \label{def_ProductCode}
The direct product $\CYCa \otimes \CYCb$ is an \LIN{\CYCan \CYCbn}{\CYCak \CYCbk}{q} code with distance $\CYCad \CYCbd$ which consists of all $\CYCan \times \CYCbn$ matrices whose rows are all codewords of $\CYCa$ and whose columns are all codewords of $\CYCb$.
\end{definition}
We recall Thm. 1 of Burton and Weldon~\cite{burton_cyclic_1965}. Throughout the paper, we restrict ourselves to the case where both codes are over the same alphabet.
\begin{theorem}[Cyclic Product Code] \label{theo_CyclicProductCode}
Let an \LIN{\CYCan}{\CYCak}{q} cyclic code $\CYCa$ with minimum distance $\CYCad$ and a second \LIN{\CYCbn}{\CYCbk}{q} cyclic code $\CYCb$ with minimum distance $\CYCbd$ be given.
The product code $\CYC = \CYCa \otimes \CYCb$ is an \LIN{\CYCan \CYCbn}{\CYCak \CYCbk}{q} cyclic code (with distance $\CYCd=\CYCad \CYCbd$) provided that the two lengths $\CYCan$ and $\CYCbn$ are relatively prime, i.e., $\inta \CYCan + \intb \CYCbn=1$ for some integers $\inta$ and $\intb$.
Let the $\CYCan \times \CYCbn$ matrix $\M{M} = \| m_{i,j}\|_{i=0,j=0}^{\CYCan-1,\CYCbn-1}$ be a codeword of the product code $\CYC$ as in Def.~\ref{def_ProductCode}. Then, the univariate polynomial $c(X) = \sum_{i=0}^{\CYCan \CYCbn -1 } c_i X^i \in \Fqx$ with
\begin{equation}
c_i =m_{i \bmod \CYCan, i \bmod \CYCbn}, \quad \forall i =0,1,\dots,\CYCan \CYCbn-1
\end{equation}
is a codeword of the cyclic product code $\CYC$ that is an ideal in the ring $\Fqx / (X^{\CYCan \CYCbn}-1)$.
\end{theorem}
Let us outline how the defining set $\defset{\CYC}$ of $\CYC = \CYCa \otimes \CYCb$ can be obtained from $\defset{\CYCa}$ and $\defset{\CYCb}$. We summarize the results of Lin and Weldon~\cite[Thm. 4]{lin_further_1970}.
\begin{theorem}[Defining Set and Generator Polynomial of a Cyclic Product Code] \label{theo_DefSetProductCode}
Let $\CYCa$ and $\CYCb$, respectively \LIN{\CYCan}{\CYCak}{q} and \LIN{\CYCbn}{\CYCbk}{q}, be cyclic codes with defining sets $\defset{\CYCa}$ and $\defset{\CYCb}$ and generator polynomials $g_1(X)$ and $g_2(X)$. Let $\inta \CYCan + \intb \CYCbn =1 $ for some integers $\inta$ and $\intb$. Then, the generator polynomial $g(X)$ of the cyclic product code $\CYCa \otimes \CYCb$ is:
\begin{equation} \label{eq_GenPolyCyclicProduct}
g(X) = \gcd \Big( X^{\CYCan \CYCbn}-1, g_1\big( X^{\intb \CYCbn} \big) \cdot g_2 \big(X^{\inta \CYCan}\big) \Big).
\end{equation}
Let $\SET{B}_{\CYCa} = \defset[\intb,\otimes]{\CYCa}$ and let $\SET{A}_{\CYCb} = \defset[\inta,\otimes]{\CYCb}$ as given in~\eqref{eq_definingsetext}. The defining set of the cyclic product code $\CYC$ is:
\begin{align*} \label{eq_DefSetCyclicProduct}
\defset{\CYC} =  \Bigg\{  \bigcup_{i=0}^{\CYCbn-1} \SET{B}_{\CYCa}^{[i \CYCan,+]} \Bigg\} \cup \Bigg\{  \bigcup_{i=0}^{\CYCan-1} \SET{A}_{\CYCb}^{[i \CYCbn,+]} \Bigg\}.
\end{align*} 
\end{theorem}
Let us restate Thm. 2 of~\cite{zeh_new_2012} on the minimum distance of cyclic codes using cyclic product codes.
\begin{theorem}[BCH Bound Generalization] \label{theo_GenBCHbound}
Let an \LIN{\CYCan}{\CYCak}{q} cyclic code $\CYCa$ with minimum distance $\CYCad$ and a second \LIN{\CYCbn}{\CYCbk}{q} cyclic code $\CYCb$ with minimum distance $\CYCbd$ and with $\gcdab{\CYCan}{\CYCbn}=1$ be given.
Let $\alpha$ be an element of order $\CYCan$ in $\F{q^{\CYCas}}$, $\beta$ of order $\CYCbn$ in $\F{q^{\CYCbs}}$ respectively. Let the integers $\HTconsta$, $\HTconstb$, $\HTmulta$, $\HTmultb$, $\lenseq$ with $\HTmulta \neq 0$, $\HTmultb \neq 0$, $\gcdab{\CYCan}{\HTmulta}=\gcdab{\CYCbn}{\HTmultb}=1$ and $\lenseq \geq 2$ be given, such that:
\begin{equation} \label{eq_statementBCHgen}
\sum_{i=0}^{\infty}  a(\alpha^{\HTconsta+i\HTmulta}) \cdot b(\beta^{\HTconstb+i\HTmultb}) X^{i} \equiv 0 \mod X^{\lenseq-1}
\end{equation}
holds for all codewords $a(X) \in \CYCa$ and $b(X) \in \CYCb$.
Then, we obtain:
\begin{equation}
\CYCad \geq d^{\ast} = \left \lceil \frac{\lenseq}{\CYCbd} \right \rceil.
\end{equation}
\end{theorem}
Note that the expression of~\eqref{eq_statementBCHgen} is in \Fxsub{q^\CYCs}, where $\CYCs = \lcmab{\CYCas}{\CYCbs}$.
\begin{proof}
From Thm.~\ref{theo_DefSetProductCode} we know that \eqref{eq_statementBCHgen} corresponds to $\lenseq-1$ consecutive zeros in the defining set $\defset{\CYC}$ of $\CYC = \CYCa \otimes \CYCb$ and therefore its distance $\CYCd =\CYCad \CYCbd$ is greater than or equal to $\lenseq$.
\end{proof}
Moreover, this yields the following explicit relation.
\begin{proposition}[BCH Bound of the Cyclic Product Code] \label{lem_ParameterBCHGenProd}
Let the integers $\HTconsta,\HTconstb,\HTmulta \neq 0,\HTmultb \neq 0$ and $\lenseq \geq 2$ and two cyclic codes $\CYCa$ and $\CYCb$ with $\inta \CYCan+ \intb \CYCbn=1$ be given as in Thm.~\ref{theo_GenBCHbound}.
Then, the two integers:
\begin{align*}
\HTconst = & \HTconsta \cdot \intb^2 \CYCbn + \HTconstb \cdot \inta^2 \CYCan \quad \quad \text{and}  \\
\HTmult = & \HTmulta \cdot \intb^2 \CYCbn + \HTmultb \cdot \inta^2 \CYCan
\end{align*}
denote the parameters such that:
\begin{equation} \label{eq_CorBCHBoundProductCode}
\sum_{i=0}^{\infty}  c(\gamma^{\HTconst+ i \HTmult  }) X^{i} \equiv 0 \mod X^{\lenseq-1}
\end{equation}
holds for all $c(X) \in \CYCa \otimes \CYCb$, where $\gamma$ is a primitive element of order $\CYCan \CYCbn$ in $\Fxsub{q^\CYCs}$.
\end{proposition}
\begin{IEEEproof}
Let $g_1(X)$ be the generator polynomial of $\CYCa$ and $g_2(X)$ that of $\CYCb$.
From Thm.~\ref{theo_DefSetProductCode} we know that if $\alpha^i$ is a root of $g_1(X)$, then $\gamma^{\intb i}$ is a root of $g(X)$ as in~\eqref{eq_GenPolyCyclicProduct} and $\gamma^{\inta i}$ is a root of $g(X)$ if $\beta^i$ is a root of $g_2(X)$.
Therefore we want $\HTconst + i \HTmult \equiv \intb (\HTconsta + i \HTmulta) \bmod \CYCan $ and $\HTconst + i \HTmult \equiv \inta (\HTconstb + i \HTmultb) \bmod \CYCbn $ and the Chinese-Remainder-Theorem gives the result.
\end{IEEEproof}

\begin{example}[BCH Bound of the Cyclic Product Code]
Let $\CYCa$ be the binary reversible $\LIN{17}{9}{2}$ code with ${\defset{\CYCa}} = \{1,2,4,8,-8,-4,-2,-1 \}$ and let $\CYCb$ denote the binary $\LIN{3}{2}{2}$ single parity check code with ${\defset{\CYCb}} = \{0\}$. Let $\alpha \in \F{2^8}$ and $\beta \in \F{2^4}$ denote elements of order 17 and 3, respectively. Then, we know that for $\HTconsta=-4$, $\HTconstb=-1$ and $\HTmulta=\HTmultb=1$ Thm.~\ref{theo_GenBCHbound} holds for $\delta=10$ and therefore $\CYCad \geq 5$, which is the true minimum distance of $\CYCa$. 

Since $-1 \cdot 17 + 6 \cdot 3 = 1$, according to Thm.~\ref{theo_DefSetProductCode} the defining set of the cyclic product code $\CYCa \otimes \CYCb$ is $\defset{\CYCa \otimes \CYCb}$:
\begin{align*}
= & \Big\{ \{ 3,5,6,7,10,11,12,14 \} \cup \{ 20,22,23,24,27,28,29,31 \} \\ 
&  \cup \{ 37,39,40,41,44,45,46,48 \} \cup \{0\} \cup \{3\} \cup \cdots \cup \{48\} \Big\} \\
= & \{ 0,3,5,6,7,9,10,11,12,14,15,18,20,21,22,23,24,27,28,\\
& \quad \quad \quad \quad 29,30,31,33,36,37,39,40,41,42,44,45,46,48 \}
\end{align*}
and Proposition~\ref{lem_ParameterBCHGenProd} gives $\HTconst = 10$ and $\HTmult=23$.
\end{example}

\section{Generalized HT Bound I: Using Cyclic Product Code} \label{sec_GenHTBoundProd}
In this section, we consider the first generalization of Thm.~\ref{theo_GenBCHbound}.
\begin{theorem}[Generalized HT Bound I] \label{theo_GenHTBoundOne}
Let an \LIN{\CYCan}{\CYCak}{q} cyclic code $\CYCa$ with minimum distance $\CYCad$ and a second \LIN{\CYCbn}{\CYCbk}{q} cyclic code $\CYCb$ with minimum distance $\CYCbd$ and $\gcdab{\CYCan}{\CYCbn}=1$ be given.
Let $\alpha$ be an element of order $\CYCan$ in $\F{q^{\CYCas}}$, $\beta$ of order $\CYCbn$ in $\F{q^{\CYCbs}}$, respectively. Let the integers $\HTconsta$, $\HTconstb$, $\HTmulta$, $\HTmultb$, $\lenseq$ and $\noseq$ with $\HTmulta \neq 0$, $\HTmultb \neq 0$, $\gcdab{\CYCan}{\HTmulta}=\gcdab{\CYCbn}{\HTmultb}=1$, $\lenseq \geq 2$ and $\noseq > 0$ be given, such that 
\begin{align} \label{eq_HTProductCodea}
\sum_{i=0}^{\infty} a(\alpha^{\HTconsta+i\HTmulta+j}) \cdot & b(\beta^{\HTconstb+i\HTmultb+j}) X^i \nonumber \\ 
& \equiv 0 \bmod X^{\lenseq-1} \quad \forall j=0,1,\dots,\noseq
\end{align}
holds for all codewords $a(X) \in \CYCa$ and  $b(X) \in \CYCb$.
Then, the minimum distance $\CYCad$ of $\CYCa$ is lower bounded by:
\begin{align} \label{eq_HTBoundProductCodeb}
\CYCad \geq d^{\ast \ast} \defeq \left \lceil \frac{\lenseq + \noseq}{\CYCbd}  \right \rceil.
\end{align}
\end{theorem}
\begin{IEEEproof}
From the generator polynomial of the cyclic product code $\CYCa \otimes \CYCb$ (see Thm.~\ref{theo_DefSetProductCode}) we know that whenever $a(X) \in \CYCa$ or $b(X) \in \CYCb$ have a zero, then a codeword of the cyclic product code $\CYCa \otimes \CYCb$ is also zero at the evaluated point (as stated in Lemma~\ref{lem_ParameterBCHGenProd}). Therefore, $\lenseq + \noseq$ is the HT bound (see Thm.~\ref{theo_HTBound}) of $\CYCa \otimes \CYCb$ and $\CYCad \CYCbd \geq \lenseq + \noseq$.
\end{IEEEproof}
\section{Generalized HT Bound II: Using a Second Cyclic Code} \label{sec_GenHTBoundFrac}
In this section, we consider the second generalization of Thm.~\ref{theo_GenBCHbound} and the proof of the statement is more involved.
\begin{theorem}[Generalized HT Bound II] \label{theo_GenHTBoundTwo}
Let an \LIN{\CYCan}{\CYCak}{q} cyclic code $\CYCa$ with minimum distance $\CYCad$ and a second \LIN{\CYCbn}{\CYCbk}{q} cyclic code $\CYCb$ with minimum distance $\CYCbd$ and with $\gcdab{\CYCan}{\CYCbn}=1$ be given.
Let $\alpha$ be a primitive element of order $\CYCan$ in $\F{q^{\CYCas}}$, $\beta$ of order $\CYCbn$ in $\F{q^{\CYCbs}}$ respectively.
Let the integers $\HTconsta$, $\HTconstb$, $\HTmulta$, $\HTmultb$, $\lenseq$ and $\noseq$ with $\HTmulta \neq 0$, $\HTmultb \neq 0$, $\gcdab{\CYCan}{\HTmulta}=\gcdab{\CYCbn}{\HTmultb}=1$, $\lenseq \geq 2$ and $\noseq > 0$ be given, such that:
\begin{align} \label{eq_HTProductCode}
\sum_{i=0}^{\infty} a(\alpha^{\HTconsta+i\HTmulta+j}) \cdot & b(\beta^{\HTconstb+i\HTmultb}) X^i \nonumber \\ 
& \equiv 0 \bmod X^{\lenseq-1} \quad \forall j=0,1,\dots,\noseq
\end{align}
holds for all codewords $a(X) \in \CYCa$ and  $b(X) \in \CYCb$. Then, the minimum distance $\CYCad$ of $\CYCa$ is lower bounded by:
\begin{align} \label{eq_HTBoundProductCode}
\CYCad \geq d^{\ast \ast \ast} \defeq \left \lceil \frac{\lenseq}{\CYCbd} + \noseq  \right \rceil.
\end{align}
\end{theorem}
\begin{IEEEproof}
Let $a(X) = \sum_{i \in \SET{Y}} a_i X^i$ with $\SET{Y} = \{i_1,i_2,\dots, i_y \}$ and $b(X) = \sum_{i \in \SET{Z}} b_i X^i$ with $\SET{Z} = \{j_1,j_2,\dots, j_{z} \}$.
We combine the $\noseq+1$ sequences (multiplying each of it by $\lambda_i \in \F{q^s},\, s=\lcmab{\CYCas}{\CYCbs}$) and obtain: 
\begin{align*} 
& \sum_{i=0}^{\infty} \Bigg( \lambda_0 \sum_{\ell \in \SET{Z}} b_{\ell}\beta^{\ell (\HTconstb + i \HTmultb)} ( a_{i_1}\alpha^{i_1(\HTconsta + i\HTmulta)} + \dots + \\ 
& a_{i_y}\alpha^{i_y(\HTconsta + i\HTmulta)} ) + \lambda_1 \sum_{\ell \in \SET{Z}} b_{\ell}\beta^{\ell (\HTconstb + i \HTmultb)}  ( a_{i_1}\alpha^{i_1(\HTconsta + i\HTmulta+1)} + \dots \\  
& + a_{i_y}\alpha^{i_y(\HTconsta + i\HTmulta+1)}) + \dots +
\lambda_{\noseq} \sum_{\ell \in \SET{Z}} b_{\ell} \beta^{\ell (\HTconstb + i \HTmultb)} \\
& ( a_{i_1}\alpha^{i_1(\HTconsta + i\HTmulta+\noseq)} + \dots + a_{i_y}\alpha^{i_y(\HTconsta + i\HTmulta+\noseq)} ) X^{i} \equiv 0 \bmod X^{\lenseq-1}.
\end{align*}
Simplified, this result in:
\begin{equation} \label{eq_HTBoundTwoExplicitIntera}
\begin{split}
 \sum_{i=0}^{\infty} b(\beta^{\HTconstb + i \HTmultb}) & \Big( \sum_{\ell \in \SET{Y}} a_{\ell} \alpha^{\ell(\HTconsta+i\HTmulta)} (\lambda_0+ \alpha^{\ell} \lambda_1 +\dots \\ 
& + \alpha^{\ell \noseq}\lambda_{\noseq}) \Big) X^i \equiv 0 \mod X^{\lenseq-1}.
\end{split}
\end{equation}
We want to annihilate the first $\noseq$ terms and guarantee that the linear combination is nonzero. The corresponding $(\noseq+1) \times (\noseq+1)$ system of equations:
\begin{align} \label{eq_SystemForCoefficients}
\begin{pmatrix}
1 & \alpha^{i_1} & \alpha^{i_1 2} & \cdots & \alpha^{i_1 \noseq}  \\ 
1 & \alpha^{i_2} & \alpha^{i_2 2} & \cdots & \alpha^{i_2 \noseq }  \\ 
 &  & \vdots &  &  \\ 
1 & \alpha^{i_{\noseq+1}} & \alpha^{i_{\noseq+1} 2} & \cdots & \alpha^{i_{\noseq+1} \noseq}  \\ 
\end{pmatrix}
\begin{pmatrix}
\lambda_0 \\ 
\lambda_1 \\ 
\vdots \\
\lambda_{\noseq}
\end{pmatrix}
= 
\begin{pmatrix}
0 \\ 
\vdots \\
0 \\
1
\end{pmatrix}.
\end{align}
has a unique nonzero solution due to full rank of the square Vandermonde matrix of order $\noseq+1$ generated
by the distinct elements $\alpha^{i_1},\alpha^{i_2},\dots,\alpha^{i_{\noseq+1}}$.

Let $\tilde{\SET{Y}} \defeq \SET{Y} \setminus \{ i_1,i_2,\dots, i_{\noseq}\}$ and~\eqref{eq_HTBoundTwoExplicitIntera} leads to:
\begin{equation*}
\begin{split}
\sum_{i=0}^{\infty} b(\beta^{\HTconstb + i \HTmultb}) & \Big( \sum_{\ell \in \tilde{\SET{Y}}} a_i \alpha^{\ell(\HTconsta+i\HTmulta)} (\lambda_0+ \alpha^{\ell} \lambda_1 +\dots+ \nonumber \\
& \alpha^{\ell \noseq} \lambda_{\noseq}) \Big) X^i  \equiv 0 \mod X^{\lenseq-1}.
\end{split}
\end{equation*}
This leads to (for the sake of clarity, we let $\HTmulta=\HTmultb=1$):
\begin{small}
\begin{align*}
& \dfrac{\sum\limits_{i \in \tilde{\SET{Y}}} \Big( a_i \alpha^{i\HTconsta} \sum\limits_{j \in \SET{Z}} \Big(b_j \beta^{j\HTconstb} \prod\limits_{\substack{\ell \in \SET{Z}\\ \ell \neq j}} (1-X\alpha^{i} \beta^{\ell}) \Big) \prod\limits_{\substack{h \in \tilde{\SET{Y}}\\ h \neq i}} \prod\limits_{p \in \SET{Z}} (1-X\alpha^{h} \beta^{p}) \Big)  }{\prod\limits_{i \in \tilde{\SET{Y}}} \big( \prod\limits_{j \in \SET{Z}} (1-X\alpha^{i} \beta^{j}) \big)} \\ 
& \equiv 0 \mod X^{\lenseq-1},
\end{align*}
\end{small}
\hspace{-.15cm}where the numerator is a nonzero linear combination of the polynomials $\prod_{(h,l) \neq (i,j)}(1-X \alpha^h \beta^l)$. It is easily shown that all of those polynomials are distinct and linearly independent (it requires that $\gcdab{\CYCan}{\CYCbn} = \gcdab{\CYCan}{\HTmulta} = \gcdab{\CYCbn}{\HTmultb} =1$). Hence, the numerator is a nonzero polynomial. Its degree is smaller than or equal to $z-1+z(y-\noseq-1) = z(y-\noseq) -1$ and therefore with $\CYCad \geq y$ and $\CYCbd \geq z$, the statement follows.
\end{IEEEproof}

\section{Decoding up to Generalized HT Bound I} \label{sec_DecGenHTBound}

Let $r(X) = a(X) + e(X)$ be the received polynomial, where $e(X) = \sum_{i \in E} e_i x^i \in \Fqx$ is the error word and $\SET{E} = \{j_1, j_2, \dots, j_t\}\subseteq \lbrace 0,\dots,\CYCan-1\rbrace$ is the set of error positions of cardinality $|\SET{E}|=t$ and $a(X)$ is a codeword of a given $\LINq{\CYCan}{\CYCak}$ code $\CYCa$.

We describe how to decode up to the generalized bound from Thm.~\ref{theo_GenHTBoundOne}. Therefore, we want to decode $t \leq \tau$ errors, where
\begin{equation}\label{eq_DecRadius}
\tau \leq \frac{d^{\ast \ast}-1}{2} = \frac{\lenseq + \noseq -1}{2 \CYCbd}.
\end{equation}
Let $b(X) \in \CYCb$ be of weight $\CYCbd$ and $\alpha \in  \F{q^{\CYCas}}$, $\beta \in  \F{q^{\CYCbs}}$ and the integers $\HTconsta, \HTconstb, \HTmulta \neq 0, \HTmultb \neq 0$ be given such that Thm.~\ref{theo_GenHTBoundOne} for $\lenseq$ and $\noseq$ holds. Denote $\CYCs =\lcmab{\CYCas}{\CYCbs}$. We define $\noseq+1$ syndrome polynomials $S^{(j)}(X) \in \Fxsub{q^s}$ for $j=0,\dots,\noseq$ as follows:
\begin{align} \label{eq_DefSyndromes}
S^{(j)}(X) & \defequiv  \sum \limits_{i=0}^{\infty} r(\alpha^{\HTconsta+i\HTmulta+j}) \cdot b(\beta^{\HTconstb+i\HTmultb+j}) X^i \bmod X^{\lenseq-1} \nonumber \\
& = \sum \limits_{i=0}^{\lenseq-2} e(\alpha^{\HTconsta+i\HTmulta+j}) \cdot b(\beta^{\HTconstb+i\HTmultb+j}) X^i.
\end{align}
This generalizes our previous approach~\cite{zeh_decoding_2012} to $\noseq+1$ syndrome sequences of length $\lenseq -1$. Hence, we obtain $\noseq+1$ key equations with a common error-locator polynomial $\Lambda(X) \in \Fxsub{q^s}$ of degree $\CYCbd t$ (compare also \cite[Equation~(20)]{zeh_decoding_2012}):
\begin{equation*}
\Omega^{(j)}(X) \equiv \Lambda(X) \cdot S^{(j)}(X)  \bmod X^{\lenseq-1}, \quad j=0,\dots,\noseq,
\end{equation*}
where the degree of $\Omega^{(j)}(X)$ is less than $\CYCbd t$.
Solving these $\noseq+1$ key equations jointly is a multi-sequence shift-register synthesis problem for sequences of equal length;
for efficient algorithms see e.g. Feng--Tzeng~\cite{feng_generalized_1989, feng_generalization_1991}.

The basic task is to solve the following linear system of equations for $\Lambda(X) = \Lambda_0 + \Lambda_1 X + \dots + \Lambda_{\CYCbd t}X^{\CYCbd t}$, which we normalized such that $\Lambda_0=1$:
\begin{equation} \label{eq_KEWithMatrices} 
\begin{pmatrix}
\mathbf{S}^{(0)}\\
\mathbf{S}^{(1)}\\
\vdots \\
\mathbf{S}^{(\noseq)}\\
\end{pmatrix}
\cdot
\begin{pmatrix}
\Lambda_{\CYCbd t}\\
\vdots \\
\Lambda_2\\
\Lambda_1\\
\end{pmatrix}
=
\begin{pmatrix}
\mathbf{T}^{(0)} \\
\mathbf{T}^{(1)} \\
\vdots \\
\mathbf{T}^{(\noseq)} \\
\end{pmatrix},
\end{equation}
where each sub-matrix $\mathbf{S}^{(j)}$ is a $(\lenseq-1-\CYCbd t) \times(\CYCbd t)$ matrix and $\mathbf{T}^{(j)}$ is a column vector of length $\lenseq-1-\CYCbd t$ as follows: 
\begin{equation} \label{eq_SubmatrixSyndromes}
\mathbf{S}^{(j)} =  \begin{pmatrix}
S_{0}^{(j)}& S_{1}^{(j)}& \dots& S_{\CYCbd t-1}^{(j)}\\
S_{1}^{(j)}& S_{2}^{(j)}& \dots& S_{\CYCbd t}^{(j)}\\
\vdots & & & \vdots \\
S_{\lenseq-2-\CYCbd t}^{(j)}& S_{\lenseq-1-\CYCbd t}^{(j)}& \dots& S_{\lenseq-3}^{(j)}
 \end{pmatrix}
\end{equation}
and $\mathbf{T}^{(j)} = ( S_{\CYCbd t}^{(j)} \, , \, S_{\CYCbd t+1}^{(j)} \, , \, \dots \, , \, S_{\lenseq-2}^{(j)} )^T$.
In the following, denote $\mathbf{S} \defeq (\mathbf{S}^{(0)T}\, , \,\mathbf{S}^{(1)T}\, , \, \dots \, , \,\mathbf{S}^{(\noseq)T})^T$. In order to guarantee unique decoding, we have to prove that the syndrome matrix $\mathbf{S}$ from \eqref{eq_KEWithMatrices} has full rank if \eqref{eq_DecRadius} is fulfilled.
For simplicity, we consider only a single parity check code for $\CYCb$ with $\CYCbd=2$.

\begin{theorem}[Decoding up to Generalized HT Bound I for a single parity check code with $\CYCbd=2$] \label{theo_DecodingGeneralizedBound}
Let $\CYCb$ be a single parity check code with $\CYCbd = 2$ and let $\gcdab{\CYCan}{\CYCbn} = \gcdab{\CYCan}{\HTmulta} = \gcdab{\CYCbn}{\HTmultb}=1$ hold.
Moreover, let \eqref{eq_DecRadius} be fulfilled and let $\noseq +1$ syndrome sequences of length $\lenseq-1$ be defined as in \eqref{eq_DefSyndromes}.
Then, the syndrome matrix $\M{S}$ with the submatrices from~\eqref{eq_SubmatrixSyndromes} has $\rank(\M{S})=2t$.
\end{theorem}

\begin{IEEEproof}
Let us w.l.o.g. assume that $b(X) = 1 + X$ and $\HTconsta = \HTconstb = 0$. Then, the $\noseq+1$ syndrome polynomials in $\Fxsub{q^s}$ are $S^{(j)}(X)  = \sum_{i=0}^{\lenseq-2} e(\alpha^{i\HTmulta + j}) (1 +\beta^{i\HTmultb+j}) X^i$ for $j=0,1,\dots,\noseq $. 
Similar to \cite[Section~VI]{feng_generalization_1991}, we can decompose the syndrome matrix into three matrices as follows.
\begin{equation*}
\M{S} = 
\begin{pmatrix}
\M{S}^{(0)}\\
\vdots\\
\M{S}^{(\noseq)}\\
\end{pmatrix}
= \M{X} \cdot \M{Y} \cdot \overline{\M{X}}
=\begin{pmatrix}
\M{X}^{(0)}\\
\vdots\\
\M{X}^{(\noseq)}\\
\end{pmatrix}
\cdot
\M{Y} \cdot \overline{\M{X}}, 
\end{equation*}
where $\M{X}$ is a $(\noseq+1)(\lenseq-1-2t) \times 2t$ matrix over $\F{q^s}$ and $\M{Y}$ and $\overline{\M{X}}$ are $2t \times 2t$ matrices over $\F{q}$ and $\F{q^s}$, respectively.
The decomposition provides the following matrices with $\kappa = \lenseq - 2 -2t$: 
\begin{align*}
&\M{X}^{(j)}= \left( \begin{array}{ccc}
\alpha^{j_1 j} & \dots & \alpha^{j_t j}  \\ 
\alpha^{j_1(j +\HTmulta)} & \dots & \alpha^{j_t(j +\HTmulta)} \\ 
\vdots & & \vdots \\
\alpha^{j_1(j +\HTmulta(\kappa))} & \dots & \alpha^{j_t(j +\HTmulta(\kappa))}
\end{array} \right. \\
& \left.
 \begin{array}{ccc}
\beta^{j } \alpha^{j_1 j}  & \dots & \beta^{j} \alpha^{j_t j} \\
\beta^{j +\HTmultb} \alpha^{j_1(j+ \HTmulta)} & \dots &\beta^{j+ \HTmultb} \alpha^{j_t(j+ \HTmulta)} \\
\vdots & & \vdots \\
\beta^{j +\HTmultb (\kappa)t} \alpha^{j_1(j+ \HTmulta(\kappa))} &\dots &\beta^{j+ \HTmultb (\kappa)} \alpha^{j_t(j + \HTmulta(\kappa))}
\end{array} \right),
\end{align*}
and $\M{Y} = \diag(e_{j_1},e_{j_2},\dots,e_{j_t},e_{j_1},e_{j_2},\dots, e_{j_t})$ and
\begin{equation*}
 \overline{\M{X}} = 
\begin{pmatrix}
1 & \alpha^{j_1 \HTmulta} & \dots & \alpha^{j_1\HTmulta (2t-1)}\\
1 & \alpha^{j_2 \HTmulta} & \dots & \alpha^{j_2\HTmulta (2t-1)}\\
& \vdots & & \vdots \\
1 & \alpha^{j_t \HTmulta} & \dots &\alpha^{j_t\HTmulta (2t-1)}\\
\hline 
1 & \beta^{\HTmultb} \alpha^{j_1 \HTmulta}  & \dots & (\beta^{\HTmultb} \alpha^{j_1 \HTmulta} )^{(2t-1)}\\
1 & \beta^{\HTmultb} \alpha^{j_2 \HTmulta}  & \dots & (\beta^{\HTmultb} \alpha^{j_2 \HTmulta} )^{(2t-1)}\\
& \vdots & & \vdots \\
1 & \beta^{\HTmultb} \alpha^{j_t \HTmulta}  & \dots & (\beta^{\HTmultb} \alpha^{j_t \HTmulta})^{(2t-1)}\\
\end{pmatrix}.
\end{equation*}
Since $\M{Y}$ is a diagonal matrix, it is non-singular.
From $\gcdab{\CYCan}{\CYCbn} = \gcdab{\CYCan}{\HTmulta} = \gcdab{\CYCbn}{\HTmultb}=1$ we know that $\overline{\M{X}}$ is a Vandermonde matrix and has full rank. 
Hence, $\M{Y} \cdot \overline{\M{X}}$ is a non-singluar $2t \times 2t$ matrix and therefore $\rank(\M{S}) = \rank(\M{X})$. 
In order to analyze the rank of $\M{X}$, we proceed similarly as in \cite[Sec.~VI]{feng_generalization_1991}.
We use the matrix operation from \cite{van_lint_minimum_1986} (see Corollary~\ref{coro_MatrixRank} in the appendix)
to rewrite $\M{X} = \M{A} * \M{B}$, where
\begin{equation*}
\M{A} = \begin{pmatrix}
1 & \dots &1 &1 &\dots &1\\
 \alpha^{j_1} & \dots & \alpha^{j_t }  & \beta\alpha^{j_1}  & \dots & \beta\alpha^{j_t}  \\
  \vdots &&&& &\vdots\\
  \alpha^{j_1 \noseq } & \dots & \alpha^{j_t \noseq }  & (\beta\alpha^{j_1})^{\noseq}  & \dots & (\beta\alpha^{j_t})^{\noseq }  \\
 \end{pmatrix}
\end{equation*}
and $\M{B} = \M{X}^{(0)}$.

Since $\gcdab{\CYCan}{\CYCbn} = \gcdab{\CYCan}{\HTmulta} = \gcdab{\CYCbn}{\HTmultb}=1$,
both matrices $\M{A}$ and $\M{B}$ are Vandermonde matrices with ranks:
\begin{equation*}
\rank(\M{A}) = \min\{\noseq+1,2t\}, \ \rank(\M{B}) = \min\{\lenseq-1-2t,2t\}.
\end{equation*}
Note that w.l.o.g. we can always define $\HTmulta,\HTmultb,\lenseq$ and $\noseq$ such that $\noseq +1 \leq \lenseq-1$. 
Therefore, from \eqref{eq_DecRadius} we obtain:
\begin{equation} \label{eq_DecRadiuForRank}
t \leq \frac{d^{\ast \ast}-1}{2} = \frac{\lenseq + \noseq -1}{2 \CYCbd} \leq \frac{2(\lenseq-1) -1}{2 \CYCbd} < \frac{\lenseq-1}{\CYCbd}.
\end{equation}
Hence, investigating all possible four cases of $\rank(\M{A}) + \rank(\M{B})$ gives:
\begin{align*}
&2t + 2t = 4t > 2t,\\
&2t + \noseq +1 > 2t, \\
&\lenseq-1-2t +2t = \lenseq -1 > 2t,\\
&\lenseq-1-2t+\noseq+1 \geq 2 \CYCbd t - 2t +1 = 2t +1 > 2t,
\end{align*}
where the last two above inequalities used~\eqref{eq_DecRadiuForRank} and $\CYCbd = 2$.
Thus, $\rank(\M{A}) + \rank(\M{B}) > 2t$.
With Corollary~\ref{coro_MatrixRank} in the appendix, we have proven the statement.
\end{IEEEproof}
Therefore, the joint key equation \eqref{eq_KEWithMatrices} has a unique solution, which can be found by multi-sequence shift-register synthesis with $\mathcal O(sn^2)$ operations over $\F{q^s}$~\cite{feng_generalized_1989, feng_generalization_1991}. 
The extension of the proof for decoding up to $t \leq \tau$ errors as in~\eqref{eq_DecRadius} to other associated codes $\CYCb$ with $\CYCbd \geq 2$ is straight-forward. The decomposition of the syndrome matrix $\M{S}$ can be done similarly and we can prove that it has rank $\CYCbd t$.
The details of the root-finding of $\Lambda(X)$ to obtain the error-locations and the determination of the error-values can be found in Sec. 6 of~\cite{zeh_new_2012}.

\section{Conclusion} \label{sec_Conclusion}
We presented two techniques to generalize the HT bound on the minimum Hamming distance of $q$-ary cyclic codes.
The first one is directly related to cyclic product codes and facilitates a syndrome-based algebraic decoding algorithm. The second approach's connection to   product codes is an open topic as well as a decoding approach up to this bound. 

Probably, it is possible to generalize other bounds (Roos, van Lint--Wilson) on the minimum distance of cyclic codes by embedding the given code into a cyclic product code. Furthermore, it seems possible to apply this approach similarly to the wider class of linear codes.


\section*{Acknowledgments}
The authors are grateful to Daniel Augot for stimulating discussions. 


\section*{Appendix}
The following corollary follows directly from Thm.~4~\cite{van_lint_minimum_1986}.
\begin{corollary}[vLW-Matrix Product and Rank] \label{coro_MatrixRank}
Let the following matrix operation be defined as in \cite{van_lint_minimum_1986}:
\begin{equation*}
\M{X}  = \M{A} * \M{B} = 
\begin{pmatrix}
a_{1,1} \b_1 & a_{1,2} \b_2 & \dots &a_{1,2t}\b_{2t}\\
a_{2,1} \b_1 & a_{2,2} \b_2 & \dots& a_{2,2t}\b_{2t}\\
\vdots & & &\vdots\\
a_{\noseq+1,1} \b_1 & a_{\noseq+1,2} \b_2 & \dots& a_{\noseq+1,2t}\b_{2t}\\
\end{pmatrix},
\end{equation*}
where $\M{A}$ is a $(\noseq +1) \times 2t$ matrix, $\M{B}$ is a $(\lenseq-1-2t) \times 2t$ matrix and $\b_i$ denotes the $i$-th column of $\M{B}$,
and $\M{X}$ has $2t$ columns. 
If $\rank(\M{A}) + \rank(\M{B}) > 2t$, then $\rank(\M{X}) = 2t$.
\end{corollary}
\printbibliography
\end{document}